\documentclass[aps,prl,twocolumn,groupedaddress]{revtex4}
\usepackage{epsf}
\usepackage{graphicx}

\input{tcilatex}

\begin{document}

\title{Mobility and its temperature dependence in LSCO: viscous motion ?}
\author{Lev P. Gor'kov}
\affiliation{NHMFL, Florida State University, 1800 E P.Dirac Dr., Tallahassee FL 32310,
USA}
\affiliation{L.D.Landau Institute for Theoretical Physics of the RAS, Chernogolovka
142432, RUSSIA }
\author{Gregory B. Teitel'baum}
\email{grteit@kfti.knc.ru}
\affiliation{E.K.Zavoiskii Institute for Technical Physics of the RAS, Kazan 420029,
RUSSIA }
\date{\today}

\begin{abstract}
We argue that charges in underdoped La$_{2-x}$Sr$_{x}$CuO$_{4}$
move in a dissipative environment of strong spatial and temporal
fluctuations. The unusual temperature dependence of the Hall angle
known as "the separation of lifetimes" is reinterpreted and
attributed to the appearance of the thermally activated component
in the effective number of carriers with the temperature increase.
We consider the temperature interval above $T_{c}$ where
localization effects can be neglected.
\end{abstract}

\maketitle


%\keywords{Proximity effect, superconducting spin valve}

The normal phase properties of high temperature superconducting
(HT$_{c}$) cuprates differ strongly from those of ordinary
superconductors. The fact has been justly attributed to the
proximity of new materials to the Mott metal-insulator (MI)
transition, where metallic features come about due to doping of
external carriers into the CuO$_2$-plane. The Mott's physics has
been best traced in evolution of two parent materials,
La$_{2}$CuO$_{4}$ (LCO, or 214) and YBa$_{2}$Cu$_{3}$O$_{6}$
(YBCO$_{6}$, or 123), from the antiferromagnetic (AFM) insulator state into the HT%
$_{c}$ material at doping. La$_{2}$CuO$_{4}$, when doped by the
divalent Sr, transforms into the single-plane HT$_{c}$
superconductor, La$_{2-x}$Sr$_{x}$CuO$_{4}$ (LSCO). YBCO$_{6+y}$
is nowadays a classic example of the two-plane HT$_c$ material
doped by the excessive oxygen.

We concentrate on transport characteristics of LSCO. Better single
crystals have been available for these materials which have been
examined extensively. Early resistivity and the Hall effect
studies were conducted usually below room temperatures (For brief
review of old results, see e.g., \cite{Timusk}). Recent data
\cite{Ono} for resistivity and the Hall effect in LSCO now cover
the broad range of concentration and up to 1000 K. Below we apply
our analysis mostly to the findings \cite{Ono}.

In this presentation we interpret anew the $T$-dependence of
resistivity taking into account the increase of number of carriers
with temperature

Transport data for cuprates could not be easy explained in terms
of the Fermi liquid (FL) theory. Thus, resistivity in optimally
doped LSCO has been found to increase linearly with temperature up
to 1000 K without any tendency to saturation \cite{Gurvitch}. This
dependence was interpreted in frameworks of the phenomenological
Marginal Fermi Liquid (MFL) theory \cite{Varma}. According to
\cite{Varma}, all measured quantities are expected to scale with
the only possible dimensional parameter -
the temperature; hence, linear in $T$ dependence of the relaxation rate: $%
1/\tau \propto T$. A challenge to MFL came up with observation of
almost quadratic temperature dependence for the Hall angle (more
precisely, of $\cot (\theta _{H})=\rho_{xx}/\rho_{yx}$, where
$\rho_{xx}$ and $\rho_{yx}$ are the
longitudinal and the transverse resistivity components, correspondingly) %
\cite{Ong}. Quadratic $T$-dependence for the Hall angle, as
opposed to the $T$-linear resistivity, seemed to be the evidence
in favor of two scales in the relaxation processes for carriers
\cite{Chien}. (The controversy is sometimes referred to as ``the
separation of lifetimes''; e.g. see in  \cite{Hussey}).

In \cite{Anderson} the puzzle was attributed to ``spin-charge
separation'', the concept borrowed from the physics of
one-dimensional conductors and merely postulated for HT$_{c}$
cuprates. The quadratic $T$- dependence for $\cot (\theta _{H})$
was ascribed in \cite{AbVa} to small angle scattering of carriers
on dopants, e. g., on the Sr$^{2+}$- ions located far enough from
the conducting CuO$_{2}$-planes.

Treatment of transport phenomena in metals and semiconductors is
based on the fundamental concept of quasiparicles and on
the subsequent use of the Boltzmann-like equation. In the MFL theory \cite%
{Varma} the energy spectrum of electronic liquid bears the
singular character and the well-defined quasiparticles are absent.
MFL scaling \cite{Varma} does not immediately include the small
angle scattering and the Boltzmann equation approach had to be
generalized in \cite{AbVa}.

The electron energy spectrum in cuprates was directly addressed in
the ARPES experiments. Consensus is that within the current
resolution well-defined quasiparticle excitations exist only at
crossing of the ``Fermi surface (FS) locus''\ along the nodal
directions. These regions of the FS are termed as ``the Fermi
arcs'', with the arcs' lengths increasing with the increase of
temperature. Broad features are seen instead
for all other directions (see a summary of recent ARPES findings in \cite%
{Yoshida} together with a discussion concerning possible implications of the
``arcs''\ to transport properties).

It is worth mention here in passing that the Fermi surface
``restoration'', i.e. formation of sharp excitations, takes place
below $T_{c}$ \cite{Kanigel}.

The text-book expression for conductivity has the form:
\begin{equation}
\sigma =ne^{2}\tau _{tr}/m^{\ast } \label{eq1}
\end{equation}%
where $m^{\ast }$ is the effective mass, $\tau _{tr}$-the
transport scattering time (Eq.(1) can be equally expressed through
the mobility: $\mu \equiv \tau /m^{\ast }$).

The Hall coefficient, $R_{H}$, in metals and semiconductors must
also be derived using the Boltzmann equation. For the parabolic
energy spectrum the well-known result is:
\begin{equation}
R_{H}=1/nec \label{eq2}
\end{equation}%
$R_{H}$ preserves its form, Eq.(2), for interacting electrons with
the isotropic energy spectrum \cite{Khodas}. In a more general
case, however, the expression for $R_{H}$ would depend on the
model. Recall that even for semiconductors with small elliptic
pockets the expression (2) should be multiplied by a factor that
depends on the anisotropy of masses. In metals even the sign of
$R_{H}$ may depend on the FS topology \cite{Ong2}.

Eq.(2) becomes exact in the limit of strong magnetic fields
\cite{LL}. At weaker fields expression (2) is nothing more than an
\emph{estimate for effective number of carriers}. To the best of
the authors' knowledge, the only example when $R_{H}$ in its form
(2) measures the exact number of carriers is given by the motion
of charged particles in electric and magnetic fields in
a viscous media. All the more is it interesting that in case of La$_{2-x}$Sr$_{x}$CuO$%
_{4}$ experimentally the number of carriers, $n$, calculated as in
Eq.(2) exactly coincides with $x$ at small $x$ \cite{Ono, LGLT,
Tsukuba}.

It is common in the literature to consider peculiarities in
transport properties of cuprates above $T_{c}$ as coming due to
the non-FL behavior, in other words, due to non-existence of
quasiparticles in a system of strongly interacting electrons.
Strong interactions are of course important in a system near the
Mott MI transition, but the view itself does not lead to
theoretical understanding. Below we suggest that anomalies in the
cuprates' transport may actually stem from some qualitatively
different physics. It concerns, first of all, homogeneity of the
electronic liquid in cuprates.

Interpretations of the electronic spectra as obtained from the
ARPES data, for instance, always implicitly infer that studied
samples are homogeneous both in space and time. It is definitely
not so. Abstracting from non-homogeneity caused by external
doping, it is now the well-established experimental fact that
spatial and temporal fluctuations between non-magnetic regions and
incommensurate antiferromagnetic (ICAFM) regions (known also as
``stripes'') constitute the
ubiquitous feature of the so-called pseudogap (PG) phase on the ($T,x$%
)-plane for LSCO. At lower temperatures the two fluctuating phases
realize themselves as static SC regions that coexist spatially
with ICAFM areas. \emph{Static} phase coexistence is established
in the elastic neutron
experiments both with \cite{Khaykovich, Lake} and without magnetic fields \cite%
{Tranquada}, from the NMR data \cite{Mitrovic} and from the $\mu
$SR experiments \cite{Savichi}. The ``granular''\ character of
LSCO samples manifests itself in the anomalous ``$\ln
T$-resistivity''\ at low
temperatures when SC is suppressed by strong enough magnetic fields \cite%
{Ando}.

\emph{Temporal} phases' fluctuations at higher temperatures were
seen by the
inelastic neutron scattering \cite{Obzor} and in the NMR experiments \cite%
{LPGT_JETP}. Slowing down of the fluctuations at cooling, for
instance, was directly traced as ``wipe out''\ of the
$^{63}$Cu-signal at the temperature decrease \cite{wipeout}.

In the complex dynamic regime of fluctuating sub-phases kinetic
properties cannot be obtained from a Boltzmann-like approach. Note
the important role of non-elastic events in such a regime.

The clue to the following analysis is this. As it was already
emphasized above, for LSCO at small $x$ the number of externally
doped holes is known \emph{a priori} and is equal to the number of
dopants, Sr$^{2+}$.
The remarkable fact is then that from Eq.(2) and the experimental $R_{H}$ %
\cite{Ono} one indeed obtains exactly $x$ carriers per Cu-site (at
small $x$ and temperatures around 100 K) \cite{Ono, LGLT}. It
encourages us to add more significance to measurements of the Hall
coefficient in LSCO. More specifically, \textit{we assume that
from expression (2) for $R_{H}$ one obtains the true number of
carriers, }$n_{eff}$\textit{(}$T,x$\textit{), at all given
}$x$\textit{\ and temperature, }$T$\textit{\ }\cite{LGLT}. In
accordance with the introductory remarks above, we conclude that
\textit{at low }$x$\textit{\ and finite T single charges move in a
dissipative media}.

Analysis of the Hall data from \cite{Ono} performed in \cite{LGLT,
Tsukuba} has shown that the number of holes per Cu atom,
$n_{Hall}(T,x) = nV_{Cu}$, in LSCO changes with temperature as:
\begin{equation}
n_{Hall}(T,x)=n_{0}(x)+n_{1}(x)\cdot \exp [-\Delta (x)/T)]
\label{eq3}
\end{equation}%
where $n_{0}(x)=x$ at low $x$; $n_{1}$ is a constant ($\sim 2.8$),
but decreases rapidly above $x\sim 0.2$, $V_{Cu}$ is the unit
volume per Cu. Note that the activation character of the
$T$-dependent term in Eq.(3) is the thermodynamic feature.

In the ARPES experiments one also measures the energy position of
\begin{figure}[tbp]
\centering \includegraphics[height=7cm]{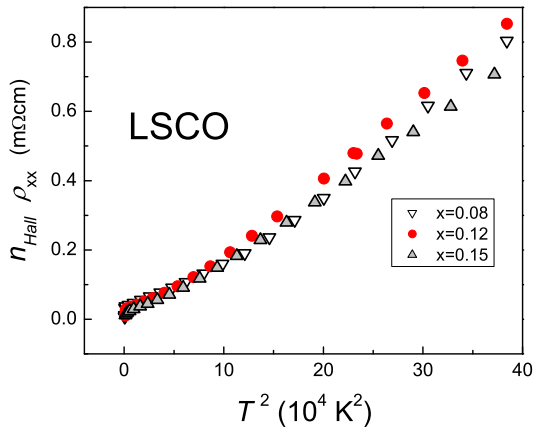} %\lable{fig:1}
\caption{ The resistivity multiplied by $n_{Hall}(T,x)$ from Eq.
(3) plotted against $T^2$ for LSCO at selected $x$ deduced from
\cite{Ono}.}
\end{figure}
the ``van Hove flat band''\ with respect to the chemical
potential. The gap, $\Delta (x)$, in (3) and this energy do
coincide, and therefore in \cite{LGLT, Tsukuba} the activation gap
has been interpreted as the ionization energy of coupled
electron-hole pairs, i.e., the local formations on Cu-O clusters.

With Eq.(3) in mind, it becomes tempting to extract the proper
relaxation rate, $1/\tau (T)$, for a single moving charge by
making use of Eq.(1). (At least, at small enough $x$ this quantity
should not depend on the holes concentration!). In Fig.1 we have
plotted (for a few
concentrations)  resistivity, $\rho (T,x)$,  multiplied by $n_{Hall}(T,x)$ \cite%
{Ono, LGLT}. One sees that all three curves at $T<$300-400 K
superimpose on each other, the result consistent with the notion
of the single charge carrier mobility.

The $T$-dependence in Fig.1 is very close to the quadratic law, $T^{2}$%
. One may try to interpret this dependence as the FL behavior of
the carriers forming small Fermi-pockets. We show that such
interpretation is not correct. Indeed, assuming a Fermi energy,
$T_{F}$, for such a hypothetical pocket, one may attempt to
re-write $\hbar /\tau $ as
\begin{equation}
\hbar /\tau =const\cdot T(m/m^{\ast })(T/T_{F})  \label{eq4}
\end{equation}%
and estimate $T_{F}$ from Fig.1. After trivial calculations one
arrives to the value: $T_{F}\sim 120-160$ K (For the effective
mass, $m^{\ast }$, we use its optical value $\sim (3-4)m_{0}$ from
\cite{Padilla}). It is obvious that the FL concept is not
applicable: the quadratic dependence on temperature in the FL
frameworks is justified only at $T<<T_{F} $.

There is no immediate explanation either for the $T$-squared
dependence that would follow out the picture of a charge moving
along fluctuating sub-phases. Most probable, the form of Eq.(4) is
nothing but a good numerical fit to the data. Note that the
quadratic dependence in Fig.1 is actually the same
$T^{2}$-dependence that has been first observed in old
measurements for the Hall angle and comes about according to (3)
after multiplying resistivity by $n_{Hall}(T,x)\propto 1/R_{H}$
\cite{Levin}. More recently, the very question whether the linear
and the quadratic $T$ -behavior in resistivity and the Hall angle,
correspondingly, are ubiquitous for cuprates became the subject of
debates (see, e.g. in \cite{Hussey}).

In Fig.2, for completeness, we plotted the same value for $\
x=0.12$ for the whole temperature interval available in
\begin{figure}[tbp]
\centering \includegraphics[height=7cm]{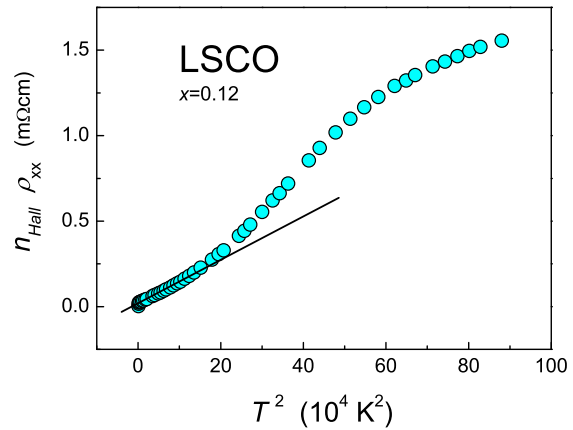} %\lable{fig:2}
\caption{The resistivity multiplied by $n_{Hall}(T,x)$  plotted
\emph{vs} $T^2$ for LSCO (x=0.12) (deduced from \cite{Ono}).}
\end{figure}
\cite{Ono}. There are two regions of a seemingly quadratic
dependence in $T$, separated by an intermediate temperature range.
The transition between the two regime occurs near the pseudogap
temperature, $T^{\ast }(x)$, for this concentration. (According to
\cite{LGLT}, $T^{\ast }(x)$ is defined as a temperature at which
the number of holes, $n_{0}(x)$, introduced through the external
doping and the number of the thermally activated holes in Eq.(3)
become approximately equal). Although at higher temperatures  the
carriers' concentration also rapidly increases and  charges may be
not independent anymore, qualitatively the result would mean that
the dissipation rate grows enormously above $\sim 100-150$ K (in
other terms, the mobility becomes extremely low).

It is only natural to wonder whether the concept elaborated above
for LSCO does apply to other HT$_{c}$ materials. Unlike LSCO, in
other cuprates there is no easy way to know the amount of holes,
$p$, introduced by the external (chemical) doping, especially
close to the onset of superconductivity. Judgments about the
actual hole concentration are then often based on the shape of the
so called ``superconducting dome'', its dependence on the dopants'
concentration, and the subsequent comparison with those in LSCO as
a template (or estimated otherwise from the thermopower experiments above $%
T_{c}$). As to the shape of the ``superconducting dome'', it is worth
reminding that for a $d$-wave SC $T_{c}$ is sensitive to defects introduced
by the doping process.

Consider briefly two examples. The low temperature Hall effect in
the normal state was measured in \cite{FedorNature} for Bi$_{2}$ Sr$_{2-x}$%
La$_{x}$CuO$_{6+y}$ (BSLCO, or La-doped Bi-2201) where the hole
number, $p$, was defined according to such a procedure in
\cite{Ando 2}. Superconductivity occurs only above $p\sim 0.10$.
We have calculated
the number of holes in La-doped Bi-2201 from the $R_{H}(T,x)$ \cite%
{FedorNature} according to Eq.(3). The results indeed turned out
very close to our results for LSCO (See Fig.3). Together with the
\begin{figure}[tbp]
\centering \includegraphics[height=7cm]{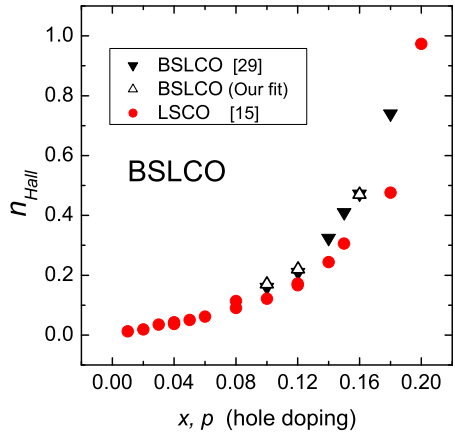} %\lable{fig:3}
\caption{The dependence of the Hall carriers number on the
concentration of the dopants  in the CuO$_2$ plane for LSCO
\cite{LGLT} (circles) and BSLCO found in \cite{FedorNature} from
$R_H$ at 100 K (filled triangles) and from the fitting of Eq. (3)
to the $R_{H}(T)$ dependencies (open triangles).}
\end{figure}
analysis \cite{LGLT} for LSCO, one may conclude that the Hall
coefficient in form of Eq.(2) does indeed serve as the measure of
the actual number of carriers. Another argument in favor of
applicability of the above physics to other materials is that
rough estimations gave us (for 0.12 - 0.15 doping) close values of
$T_{F}\sim 50$($m$*/$m$) for LSCO, YBCO \cite{YBCO} and BiSrLaCuO.
Correspondingly, the Hall angle (recalculated for the equal field
values) is practically the same for these compounds.(Noting that
this $\cot (\theta _{H})\approx AT^{2}$, it is straightforward to
obtain the expression for characteristic temperature as
$T_{F}=k_{B}m^{\ast }c/(Ae\hbar B_{z})$. Here parameter $A$ is
inversely proportional to $B_{z}$, the magnetic field normal to
CuO$_{2}$ plane. The material dependent quantities here are $A$
and $m^{\ast }$).

To summarize, according to \cite{Ono, LGLT}, the number of
carriers in SC cuprates (above $T_{c}$) changes with temperature
and deviates from simple proportionality to the amount of holes
given by concentration of chemical dopants. Therefore the
effective mobility (or $\hbar /\tau (T)$) must be calculated from
Eq.(1) after taking $n_{Hall}(T,x)$ from (3) into account. No
traditional approach is then able to explain the quadratic
\emph{T}-dependence which we consider as a purely numerical
artifact. Mobility shows dramatic decrease at temperatures above
$\sim 100$ K. We ascribe this behavior to motion of charges in a
viscous media. For cuprates spatial and temporal competition
between the two phases is ubiquitous at these temperatures.
\begin{acknowledgements}

The work of L.P.G. was supported by the NHMFL through NSF
cooperative agreement DMR-9527035 and the State of Florida, that
of  G.B.T. through the RFBR Grant N 07-02-01184.
\end{acknowledgements}

\bigskip

%\textbf{FIGURE CAPTIONS}


\begin{references}

\bibitem{Timusk} T. Timusk and B. Statt, Rep. Prog. Phys. \textbf{62},
61 (1999)

\bibitem{Ono} S. Ono, S. Komiya, and Y. Ando, Phys. Rev. B \textbf{75},
024515 (2007)

\bibitem{Gurvitch} M. Gurvitch and A. T. Fiory, Phys. Rev. Lett.
\textbf{59}, 1337 (1987)

\bibitem{Varma} C. M. Varma \emph{et al.}, Phys. Rev. Lett. \textbf{63},
1996 (1989)

\bibitem{Ong} N. P. Ong, in \emph{Physical Properties of High
Temperature Superconductors}, edited by D.M. Ginsberg, Vol. 2, p.
459 (World Scientific, Singapore, 1990)

\bibitem{Chien} T. R. Chien, Z. Z. Wang, and N. P. Ong, Phys.
Rev. Lett. \textbf{67}, 2088 (1991)

\bibitem{Hussey} N. E. Hussey, in \emph{Handbook of High Temperature
Superconductivity: Theory and Experiment} (eds. J.R. Schrieffer
and J. Brooks)(Springer Verlag, Amsterdam, 2007)

\bibitem{Anderson} P. W. Anderson, Phys. Rev. Lett. \textbf{67}, 2092
(1991)

\bibitem{AbVa} Elihu Abrahams and C. M. Varma, Phys. Rev. B \textbf{68},
094502 (2003); C.M. Varma and Elihu Abrahams, Phys. Rev. Lett.
\textbf{86}, 4652 (2001)

\bibitem{Yoshida} T. Yoshida \emph{et al.}, J. Phys.: Condens. Matter
\textbf{19}, 125209 (2007)

\bibitem{Kanigel} A. Kanigel \emph{et al.}, Phys.
Rev. Lett. \textbf{99}, 157001 (2007)

\bibitem{Khodas} M. A. Khodas and A. M. Finkel'stein, Phys.
Rev. B \textbf{68}, 155114 (2003)

\bibitem{Ong2} N. P. Ong, Phys. Rev. B \textbf{43}, 193 (1991)

\bibitem{LL} E.M.Lifshitz and L.P.Pitaevskii, \emph{Landau and
Lifshitz Course of Theoretical physics, v.10: Physical Kinetics}
(Butterworth-Heinemann,Oxford, 1997)

\bibitem{LGLT} L.P. Gor'kov and G.B. Teitel'baum, Phys. Rev.
Lett. \textbf{97}, 247003 (2006)

\bibitem{Tsukuba} L.P. Gor'kov and G.B. Teitel'baum,
arXiv:0801.1728

\bibitem{Khaykovich} B. Khaykovich \emph{et al.}, Phys Rev B \textbf{71}, 220508 (2005)

\bibitem{Lake} B. Lake \emph{et al.}, Nature (London) \textbf{415}, 299 (2002).

%\bibitem{Lake2} B. Lake, G. Aeppli, K. N. Clausen, D. F. McMorrow, K. Lefmann,
%N. E. Hussey, N. Mangkorntong, M. Nohara, H. Takagi, T. E. Mason,
%A.Schroder, Science \textbf{291}, 1759 (2001).Mozno opustit'

\bibitem{Tranquada}  J. M. Tranquada \emph{et al.}, Nature (London) \textbf{375}, 561
(1995)

\bibitem{Mitrovic} V. F. Mitrovic \emph{et al.}, Nature (London) \textbf{413},501 (2001)

\bibitem{Savichi} A. T. Savici \emph{et al.},  Phys.Rev. B \textbf{66}, 014524 (2002)

\bibitem{Ando} Yoichi Ando \emph{et al.}, Phys. Rev. Lett. \textbf{75},
4662 (1995); I. S. Beloborodov \emph{et al.}, Phys. Rev. Lett.
\textbf{91}, 246801(4) (2003)

\bibitem{Obzor} Robert J. Birgeneau \emph{et al.},  J. Phys. Soc. Jpn. \textbf{75}, 111003 (2006)

\bibitem{LPGT_JETP} L. P. Gor'kov and G. B. Teitel'baum, JETP
Lett. \textbf{80}, 195 (2004).

\bibitem{wipeout} A. W. Hunt \emph{et al.}, Phys. Rev. Lett.
82, 4300 (1999);  G.B. Teitel'baum \emph{et al.}, Phys.Rev. B
\textbf{63}, 020507 (R) (2001)

\bibitem{Padilla} W. J. Padilla \emph{et al.}, Phys. Rev. B \textbf{72}, 060511 (2005)

\bibitem{Levin} For a better fit of experimental data it
was once assumed in \cite{Levin1} that there may be other carriers
present with the temperature dependent concentration and different
relaxation times

\bibitem{Levin1} G. A. Levin and K.F. Quader, Phys. Rev. B
\textbf{62}, 11879 (2000)

\bibitem{FedorNature} Fedor F. Balakirev \emph{et al.},
Nature (London) \textbf{424}, 912 (2003).

\bibitem{Ando 2} Y. Ando \emph{et al.}, Phys. Rev. B \textbf{61}, 14956
(2000)

\bibitem{YBCO} Kouji Segawa and Yoichi Ando, Phys. Rev. B \textbf{69}, 104521
(2004)

% arXiv:0711.4213 [ps, pdf, other]

%Title: Implication of the Mott-limit violation in high-Tc cuprates

%Authors: Yoichi Ando (Osaka University)


\end{references}
\end{document}